\documentclass[runningheads]{llncs}
\usepackage[T1]{fontenc}
\usepackage{graphicx}
\usepackage{amsmath,amssymb,amsfonts}
\usepackage{bm}
\usepackage{lipsum}
\usepackage{subcaption}
\usepackage{tabularx}
\newcolumntype{Y}{>{\centering\arraybackslash}X}

\usepackage{booktabs}
\usepackage{multirow}
\usepackage{adjustbox}
\usepackage[table,xcdraw]{xcolor}
\usepackage{hyperref}

\makeatletter

\DeclareMathOperator{\E}{\mathbb{E}}
\makeatother

\begin{document}
\title{Prior Density Learning in Variational Bayesian Phylogenetic Parameters Inference\thanks{\small Supported by NSERC, FRQNT, Genome Canada and The Digital Research Alliance of Canada.}}
\titlerunning{Prior Density Learning in VBPPI}

%
\author{Amine M. Remita\inst{1} \orcidID{0000-0003-4471-3195}\and\\ Golrokh Vitae\inst{2}\orcidID{0000-0001-7399-5473}\and\\ Abdoulaye Baniré Diallo\inst{1,2} \orcidID{0000-0002-1168-9371}}
\authorrunning{Remita \textit{et al}.}
%
\institute{Department of Computer Science \and CERMO-FC Center, Department of Biological Sciences \\ Universit\'e du Qu\'ebec \`a Montr\'eal\\
\email{remita.amine@courrier.uqam.ca}, 
\email{diallo.abdoulaye@uqam.ca}
}

\maketitle              

\begin{abstract}
The advances in variational inference are providing promising paths in Bayesian estimation problems. 
These advances make variational phylogenetic inference an alternative approach to Markov Chain Monte Carlo methods for approximating the phylogenetic posterior. 
However, one of the main drawbacks of such approaches is modelling the prior through fixed distributions, which could bias the posterior approximation if they are distant from the current data distribution.
In this paper, we propose an approach and an implementation framework to relax the rigidity of the prior densities by learning their parameters using a gradient-based method and a neural network-based parameterization. 
We applied this approach for branch lengths and evolutionary parameters estimation under several Markov chain substitution models.
The results of performed simulations show that the approach is powerful in estimating branch lengths and evolutionary model parameters. They also show that a flexible prior model could provide better results than a predefined prior model. Finally, the results highlight that using neural networks improves the initialization of the optimization of the prior density parameters. 

\end{abstract}
\section{Introduction}
The Bayesian phylogenetic community is exploring faster and more scalable alternatives to the Markov chain Monte Carlo (MCMC) approach to approximate a high dimensional Bayesian posterior \cite{Fisher2022}.
The search for other substitutes is motivated by the falling computational costs, increasing challenges in large-scale data analysis, advances in inference algorithms and implementation of efficient computational frameworks.
Some of the alternatives, reviewed in \cite{Fisher2022}, are adaptive MCMC, Hamiltonian Monte Carlo, sequential Monte Carlo and variational inference (VI).
Until recently, few studies were interested in applying classical variational inference approaches in probabilistic phylogenetic models \cite{Jojic2004,Wexler2007,Cohn2010}. 
However, VI started to gain some attraction from the phylogenetic community taking advantage of advances that made this approach more scalable, generic and accurate \cite{Zhang2019advences}, such as stochastic and black box VI algorithms \cite{Hoffman2013,Ranganath2014}, latent-variable reparametrization  \cite{Kingma2014,Rezende2014}, and probabilistic programming \cite{Carpenter2017}.
These advancements allowed designing powerful and fast variational-based algorithms to infer complex phylogenetic models \cite{Fourment2019,Dang2019,Zhang2022} and analyze large-scale phylodynamic data \cite{Ki2022}.
Few studies have evaluated variational phylogenetic models using simple settings and scenarios to acquire some understanding of the variational inference approach.
For example, some analyses assumed a fixed tree topology to estimate continuous parameters (such as branch lengths and evolutionary model parameters) and approximate their posterior \cite{Fourment2019} or marginal likelihood \cite{Fourment2020}.
Zhang and Matsen \cite{Zhang2019,Zhang2020} developed and tested a fully variational phylogenetic method that jointly infers unrooted tree topologies and their branch lengths under the JC69 substitution model \cite{Jc1969}.

Bayesian methods incorporate the practitioner's prior knowledge about the likelihood parameters through the prior distributions.
Defining an appropriate and realistic prior is a difficult task, especially in the case of small data regimes, similar sequences or parameters with complex correlations \cite{Huelsenbeck2002,Huelsenbeck2005,Nascimento2017}.
It is important to note that the variational phylogenetic methods assign fixed prior distributions with default hyperparameters to the likelihood parameters, which is a similar practice in MCMC methods \cite{Nascimento2017}.
For example, i.i.d. exponential priors with a mean of 0.1 are usually placed on the branch lengths \cite{Zhang2019,Fourment2019,Fourment2020}.
However, such a choice could bias the posterior approximation and induce high posterior probabilities in cases where the data are weak or the actual parameter values do not fall within the range specified by the priors \cite{Huelsenbeck2002,Yang2005a,Nascimento2017,Fabreti2022}.
The effect of branch length priors on Bayesian inference using MCMC has been widely investigated \cite{Yang2005a,Kolaczkowski2007}, and explanations and remedies have been proposed \cite{Rannala2012,Zhang2012a,Nelson2015}.
Therefore, it is crucial to study the impact of the prior (mis)specification on the convergence and the quality of the variational approximation of the posterior and to propose solutions to overcome this problem.

Here, we show that variational phylogenetic inference can also suffer from misspecified priors on branch lengths and less severely on sequence evolutionary parameters.
We adopt a different strategy from MCMC to improve the variational posterior approximation by leveraging the structure of the variational objective function and making the prior densities more flexible.
To do that, we propose a variational approach to relax the rigidity of the prior densities by jointly learning the prior and variational parameters using a gradient-based method and a neural network-based parameterization.
Moreover, we implemented a variational Bayesian phylogenetic framework (\href{https://github.com/maremita/nnTreeVB}{\texttt{nnTreeVB}}) to evaluate its performance, consistency and behaviour in estimating the branch lengths and evolutionary model parameters using simulated datasets.

\section{Background}
\label{sect:background}
\subsection{Notation}
Suppose \(\mathbf{X}\), the observed data, is an alignment of \(M\) nucleotide sequences with length \(N\), where \(\mathbf{X} \in \mathcal{A}^{M \times N}\) and \(\mathcal{A} = \{A, G, C, T\}\). 
We assume that each site (column \(X_n\)) in the alignment evolves independently following an alignment-amortized, unrooted and binary tree \(\tau\). \(\tau\) is associated with a vector \(\mathbf{b}\) of unobserved \(2M-3\) branch lengths.
Each branch length represents the number of hidden substitutions that happen over time from a parent node to a child node. 
To estimate the branch length vector, we assume that the process of evolution (substitutions) follows a continuous-time Markov chain model parameterized by \(\psi = \{\rho\), \(\pi\}\), where \(\rho\) is the set of relative substitution rates and \(\pi\) are the relative frequencies of the four nucleotides. 
We restrict the sum of each set of parameters, \(\rho\) and \(\pi\), to one. 
We use time-reversible Markov chain models, assuming the number of changes from one nucleotide to another is the same in both ways.
Several substitution models could be defined depending on the constraints placed on their parameters \(\psi\).
JC69 is the basic model with equal substitution rates and uniform relative frequencies \cite{Jc1969}, so \(\psi = \emptyset\).
The general time-reversible (GTR) model sets free all the parameters \(\psi\) \cite{Tavare1986,Yang1994}.
In this study, we are interested in estimating the optimal branch length vector \(\mathbf{b}\) and the model parameters \(\bm{\psi}\) given the alignment \(\mathbf{X}\) and a fixed tree topology \(\tau\).

\subsection{Bayesian phylogenetic parameter inference}
The Bayesian approach to the phylogenetic parameter estimation is based on the evaluation of the joint conditional density \(p(\mathbf{b}, \bm{\psi} \,|\,\mathbf{X}, \tau)\), which we call it the phylogenetic parameter posterior.
Using Bayes' theorem, the joint posterior is computed as
\[
    p(\mathbf{b}, \bm{\psi} \,|\,\mathbf{X}, \tau) = \frac{p(\mathbf{X}, \mathbf{b}, \bm{\psi} \,|\, \tau)}{p(\mathbf{X}\,|\,\tau)} = \frac{p(\mathbf{X}\,|\, \mathbf{b}, \bm{\psi}, \tau)\, p(\mathbf{b}, \bm{\psi})}{\iint p(\mathbf{X}\,|\, \mathbf{b}, \bm{\psi}, \tau)\, p(\mathbf{b}, \bm{\psi}) \,\mathrm{d}\mathbf{b} \,\mathrm{d}\bm{\psi} }.
\]
The Bayes' equation exposes the tree likelihood \(p(\mathbf{X}\,|\,\mathbf{b}, \bm{\psi}, \tau)\), the joint prior density \(p(\mathbf{b}, \bm{\psi})\) and the model evidence \(p(\mathbf{X}\,|\,\tau)\).
The tree likelihood is the conditional probability of the observed data given the phylogenetic parameters and the tree topology. 
It is efficiently computed using Felsenstein's pruning algorithm \cite{Felsenstein1981,Posada_2021}. The algorithm assumes the independence between the alignment sites, thus \(p(\mathbf{X}\,|\,\mathbf{b}, \bm{\psi}, \tau) = \prod_{n=1}^{N} p(\mathbf{X}_n\,|\,\mathbf{b}, \bm{\psi}, \tau)\).
The rescaling algorithm \cite{Yang2000,Ayres2019} can be used to avoid cases of underflow often occurring in large trees.
Usually, we model the joint prior density using independent distributions with fixed hyperparameters for each phylogenetic parameter. Ergo, it factorizes into \(p(\mathbf{b}, \bm{\psi}) = p(\mathbf{b})\,p(\bm{\psi}) = \prod_{m=1}^{2M-3} p(\mathbf{b}_m)\; p(\bm{\rho})\, p(\bm{\pi})\).
Finally, the model evidence is the marginal probability of the data given the tree topology that integrates over the possible values of the phylogenetic parameters. The integrals over high dimensional variables make the evidence intractable, hindering the computation of the phylogenetic parameter posterior density.

\subsection{Variational inference and parameterization}
A strategy to approximate the intractable posterior density \(p(\mathbf{b}, \bm{\psi} \,|\,\mathbf{X}, \tau)\) is leveraging the variational inference (VI) approach \cite{Jordan1999,Bishop2006}, which redefines the inference as an optimization problem.
First, VI defines a family of tractable (simple to compute) densities \(q\) parameterized by \(\phi\). 
Second, it finds among these densities the closest member \(q_{\phi}^{\ast}\) to the posterior density minimizing their Kullback–Leibler divergence \(\texttt{KL}(q_{\phi}(\mathbf{b}, \bm{\psi}) \parallel p(\mathbf{b}, \bm{\psi} \,|\,\mathbf{X}, \tau))\) by tuning the variational parameters \(\phi\).
However, computing the \texttt{KL} divergence is intractable because it involves the true posterior.
Therefore, the VI approach maximizes another objective function, the evidence lower bound (\texttt{ELBO}), which is equivalent to minimizing the \texttt{KL} divergence and does not require the computation of the intractable evidence. The equation of the \texttt{ELBO} is
\begin{equation*}
    \label{eq_elbo1}
        \texttt{ELBO}(\phi, \mathbf{X}, \tau) = \E_{q_{\phi}(\mathbf{b}, \bm{\psi})} \left[ \log \left( \frac{p(\mathbf{X} \,|\,\mathbf{b}, \bm{\psi}, \tau) \, p(\mathbf{b}, \bm{\psi})}{q_{\phi}(\mathbf{b}, \bm{\psi})} \right) \right] \leq \log p(\mathbf{X}\,|\,\tau),
\end{equation*}
which can also be written as
\begin{equation}
    \label{eq_elbo2}
        \texttt{ELBO}(\phi, \mathbf{X}, \tau) = \E_{q_{\phi}(\mathbf{b}, \bm{\psi})} \left[\log p(\mathbf{X} \,|\,\mathbf{b}, \bm{\psi}, \tau) \right] - \texttt{KL} \left(q_{\phi}(\mathbf{b}, \bm{\psi}) \parallel p(\mathbf{b}, \bm{\psi}) \right).
\end{equation}
Since the tree likelihood function is non-exponential, the expectations are computed using Monte Carlo sampling of the joint approximate posterior (joint variational density) \(q_{\phi}(\mathbf{b}, \bm{\psi})\).

Following previous variational phylogenetics studies \cite{Fourment2019,Fourment2020}, we use a Gaussian mean-field variational distribution to model the joint variational density.
In this class of variational distributions, each phylogenetic parameter \(i\) is independent and follows a Gaussian (normal) distribution defined by its distinct variational parameters \(\phi_i = \{\mu_i, \sigma_i\}\). 
Hence, the joint variational density factorizes into
\(q_{\phi}(\mathbf{b}, \bm{\psi}) = \prod_{m=1}^{2M-3} q_{\phi_{\mathbf{b}_m}}(\mathbf{b}_m)\; q_{\phi_{\bm{\rho}}}(\bm{\rho})\,q_{\phi_{\bm{\pi}}}(\bm{\pi})\).
We apply invertible transformations on the variational densities and adjust their probabilities (using tractable Jacobian determinants) to accommodate the constraints on the phylogenetic parameters (branch lengths must be non-negative, and relative rates and frequencies, each must have a sum of one).
Finally, we use a stochastic gradient ascent algorithm \cite{Kingma2015} and reparameterization gradients \cite{Kingma2014,Rezende2014} to optimize the variational parameters \(\phi\).

\section{New approach}
\label{sect:approach}
\subsection{Gradient-based learning prior density}
We notice from equation \ref{eq_elbo2} that maximizing the \texttt{ELBO} induces a regularized maximization of the tree likelihood. 
The regularization of the likelihood is maintained by the minimization of \texttt{KL} divergence term, which encourages the joint variational \(q_{\phi}(\mathbf{b}, \bm{\psi})\) and the joint prior \(p(\mathbf{b}, \bm{\psi})\) densities to be close \cite{Kingma2014,Kingma2019}.
Recall that the optimization is performed with respect to the variational parameters \(\phi\), and the parameters of the prior distributions are fixed initially.  
Therefore, minimizing the \texttt{KL} divergence in equation \ref{eq_elbo2} squeezes and drives the approximate posterior density towards the joint prior density.
The \texttt{KL} divergence could dominate and lead to underfitting the variational model if the data is weak and inconsistent \cite{Hoffman2016,Krishnan2018}.
We seek to counterbalance the regularization effect by relaxing the inflexibility of the joint prior distribution. 
To achieve this, we implement adaptable parameters \(\theta\) instead of using fixed parameters for the prior densities.
The prior parameters \(\theta\) will be learned (updated) jointly with the variational parameters \(\phi\) during the optimization of the \texttt{ELBO} using gradient ascent. 
Though the \texttt{ELBO} is now maximized with respect to \(\phi\) and \(\theta\), learning \(\theta\) does not need a reparameterization gradient because the expectation remains computed using Monte Carlo sampling of the joint variational density:
\begin{equation}
    \label{eq_elbo3}
        \texttt{ELBO}(\phi, \theta, \mathbf{X}, \tau) = \E_{q_{\phi}(\mathbf{b}, \bm{\psi})} \left[\log p(\mathbf{X} \,|\,\mathbf{b}, \bm{\psi}, \tau) \right] - \texttt{KL} \left(q_{\phi}(\mathbf{b}, \bm{\psi}) \parallel p_{\theta}(\mathbf{b}, \bm{\psi}) \right).
\end{equation}

As previously mentioned, we apply independent prior densities on each phylogenetic parameter, including the branch lengths and the substitution model parameters. 
We assign on the branch lengths independent exponential distributions with rates \(\bm{\lambda}_b\).
The JC69 model has no free parameters.
Thus, the set of prior parameters of this model is \(\theta_{JC69} = \{\bm{\lambda}_b\}\).
In the case of the GTR model, we apply Dirichlet distributions on relative substitution rates and relative frequencies with concentrations \(\bm{\alpha}_{\rho}\) and \(\bm{\alpha}_{\pi}\), respectively, so
\(\theta_{GTR} = \{\bm{\lambda}_b, \bm{\alpha}_{\rho}, \bm{\alpha}_{\pi} \}\).

\subsection{Neural network-based prior parameterization}
The new prior parameters \(\theta\) are initialized independently using fixed values or sampled values from a predefined distribution (e.g.\ uniform or normal). 
To add more flexibility to the prior density, we use differentiable feed-forward neural networks to generate the prior parameter values instead of relying on a direct gradient-based update.
A neural network (NeuralNet) is constituted of a stack of \(L\) layers of neurons, where each neuron is defined by adaptable weight vector \(\mathbf{w}\) and bias \(a\).
A layer \(l\) in a neural network with a vector input \(\zeta_{l}\) generates \(g_{l}(\mathbf{W}_{l} \zeta_{l}^{\intercal} + a_{l})\), where \(g_{l}\) is the identity function or a nonlinear real-valued activation function (e.g.\ ReLU, Softplus and Softmax).
Therefore, the prior parameters to be learned \(\theta\) are constituted of the set of weights and biases of the neural networks instead of the parameters of the distributions.

The rate vector of the branch length prior densities is produced by a neural network with an input of a uniformly-sampled random noise \(\zeta_{\bm{\lambda}_b}\):
\[
        \bm{\lambda}_b = \left[\lambda_{b_1}, \lambda_{b_2}, \ldots, \lambda_{b_{2M-3}} \right]^{\intercal} = \textrm{NeuralNet}_{\theta_{\bm{\lambda}_b}}(\zeta_{\bm{\lambda}_b}).
\]
We use independent neural networks to produce the vectors of concentrations for the relative substitution rates and the relative frequencies of the GTR model:
\[
        \bm{\alpha}_{\rho} = \left[\alpha_{{\rho}_1}, \alpha_{{\rho}_2}, \ldots, \alpha_{{\rho}_6} \right]^{\intercal} = \textrm{NeuralNet}_{\theta_{\bm{\alpha}_{\rho}}}(\zeta_{\bm{\alpha}_{\rho}}),
\]
\[
        \bm{\alpha}_{\pi} = \left[\alpha_{{\pi}_1}, \alpha_{{\pi}_2}, \alpha_{{\pi}_3}, \alpha_{{\pi}_4} \right]^{\intercal} = \textrm{NeuralNet}_{\theta_{\bm{\alpha}_{\pi}}}(\zeta_{\bm{\alpha}_{\pi}}),
\]
where \(\zeta_{\bm{\alpha}_{\rho}}\) and \(\zeta_{\bm{\alpha}_{\pi}}\) are independent random noises.

\section{Experimental Study}
We conceived \href{https://github.com/maremita/nnTreeVB}{\texttt{nnTreeVB}}, a variational Bayesian phylogenetic framework, to evaluate and compare the consistency, effectiveness and behaviour of the proposed variational phylogenetic models implemented with different prior density schemes.
The framework allows us to assess the variational models in estimating one phylogenetic parameter at a time.
Thus, \texttt{nnTreeVB} simulates multiple datasets varying the prior distribution of the considered phylogenetic parameter and drawing the remaining parameters from their default priors.
It fits the variational models implemented either with fixed or adaptable prior distributions on these datasets. 
The fixed priors using default hyperparameters can match one of the priors used in data generation.
The following sections describe the procedure used in the data simulation and the settings of the variatinoal inference models.

\subsection{Dataset simulation procedure}
A dataset comprises of a tree (associated with its branch length vector) and a sequence alignment simulated using prior distributions applied over their parameters.
Given a number of taxa \(M\), we build each tree topology by sampling from a uniform distribution to have an equal prior probability. 
The branch lengths are sampled independently from an exponential distribution with a predefined rate \(\lambda\). 
The rate is inversely proportional to the expected number of substitutions along a branch (\(\lambda = 1/\E[\mathbf{b}_m]\)). 
Afterwards, we select a substitution model and draw its parameters (if it has any) from their prior distributions. 
We assumed Dirichlet distribution for the substitution rates and relative frequencies and gamma distribution for positive parameters like the transition/transversion rate ratio \(\kappa\) in K80 \cite{Kimura1980} and HKY85 \cite{Hasegawa1985} models.

Eventually, using the tree and the substitution model and given a sequence length \(N\), we simulate the evolution of a sequence alignment based on a site-wise homogeneity strategy \cite{Spielman2015}.
This strategy evolves sequences from a root sequence (sampled from the distribution of relative frequencies) with the same substitution model over lineages and with the same set of branch lengths for alignment sites.
We simulated datasets with 16 and 64 taxa and 1000 and 5000 base pairs (bp) to investigate the impact of the data regime in terms of the tree size and the sequence alignment length, respectively.
Finally, we produced 100 replicates of each dataset characterized by a prior setting and a data size condition.

\subsection{Variational inference settings}
Using the \texttt{nnTreeVB} framework, we implemented three variational phylogenetic (\texttt{VP}) models that differ in their prior distribution computation and initialization.
The first model is the baseline, which uses fixed prior distributions (\texttt{VP-FPD}) with default hyperparameters.
The two other models follow the proposed approach described in section \ref{sect:approach}, which defines adaptable prior parameters to be learned jointly along the variational parameters while optimizing the \texttt{ELBO}.
Initializing the prior parameters with \(\mathrm{i.i.d.}\) samples from a uniform distribution corresponds to the simple \texttt{VP} model with learnable prior distributions (\texttt{VP-LPD}).
The last model (\texttt{VP-NPD}) implements feed-forward NeuralNets to generate the prior parameters from a uniform sampled input. 
Based on preliminary hyperparameters fine-tuning results, we used one hidden layer of size 32 and a ReLU activation, \(\max(0, x)\), for each NeuralNet. 
To ensure the positiveness of a parameter (like rates and concentrations), we apply a softplus function, \(\log (1 + e^x)\), on the output layer.

\texttt{nnTreeVB} implements automatic differentiation via PyTorch library \cite{Pytroch} in order to estimate the gradient of the \texttt{ELBO} and optimize it.
We evaluate the gradient using the Monte Carlo integration by sampling a single data point from the approximate density at each iteration.
The variational parameters and adaptable prior parameters are updated using the Adam algorithm \cite{Kingma2015} with a learning rate value of \(0.1\) for simple parameters and \(0.01\) for NeuralNet-generated parameters.
The training of each \texttt{VP} model is performed over 2000 iterations and replicated ten times to fit a given dataset.
In the end, we estimate each phylogenetic parameter by sampling 1000 data points from the approximate density of each model replicate.

\subsection{Performance metrics}
We investigate the convergence and the performance of the variational phylogenetic models through the two components of the \texttt{ELBO} (as in equations \ref{eq_elbo2} and \ref{eq_elbo3}): the log likelihood (\texttt{LogL}) and the \texttt{KL} divergence between the approximate and the prior densities (\texttt{KL-qprior}).
Moreover, to assess the accuracy of estimating a set of phylogenetic parameters (branch lengths, substitution rates or relative frequencies), we compute the scaled Euclidean distance (\texttt{Dist}) and the Pearson correlation coefficient (\texttt{Corr}) between the estimated vector and the actual one used in the simulation of the sequence alignment.
The scaled Euclidean distance is a transformed distance to be in the range \([0, 1]\) calculated as \(\texttt{Dist} = \frac{d}{1+d}\), where \(d\) is the Euclidean distance.
We also report the tree length (\texttt{TL}) for the evaluation of estimating the branch lengths, which is the sum of their vector.
Finally, we used the nonparametric Kruskal-Wallis H test \cite{kruskal1952} to evaluate for each performance metric the differences between the results of the models.

\section{Results}
In this section, we present the results of simulation-based experiments we performed to evaluate the performance of the two proposed \texttt{VP} models that learn the prior density parameters, \texttt{VP-LPD} and \texttt{VP-NPD}, and compare them to \texttt{VP-FPD}, which is the default model and baseline used in variational phylogenetic inference.
We demonstrate that when a phylogenetic parameter deviates from the default prior distribution, \texttt{VP-L/NPD} models are more efficient in approximating the posterior density.
Moreover, we show how using neural network prior parameterization, implemented in \texttt{VP-NPD}, improves the accuracy of the substitution model parameters estimation.

We employed the \texttt{VP} models to estimate the branch lengths and substitution model parameters from datasets simulated with different sequence alignment lengths (1000 and 5000 bp) and tree sizes (16 and 64 taxa). 
However, along the section, we highlight the results from more challenging datasets with shorter sequences (1000 bp) and larger trees (64 taxa). 
For each analysis, we fit ten times the three \texttt{VP} models on 100 dataset replicates, sample 1000 estimates from their approximate densities and report their performance metrics' averages and standard deviations.

\subsection{Branch lengths estimation performance}
\begin{table}[b]
\centering
\caption{\small \textbf{Performance of branch lengths (BL) estimation on datasets simulated with different prior means}. The datasets have 64 sequences of length 1000 bp and are simulated with JC69 substitution model. BL values are drawn from an exponential distribution. \texttt{VP} models implement a JC69 model and apply an i.i.d. exponential prior on BL. \texttt{VP-FPD} applies a prior with mean 0.1.}
\begin{adjustbox}{width=1\textwidth}

\begin{tabular}{@{}c
>{\columncolor[HTML]{EFEFEF}}c 
>{\columncolor[HTML]{EFEFEF}}c 
>{\columncolor[HTML]{EFEFEF}}c ccc
>{\columncolor[HTML]{EFEFEF}}c 
>{\columncolor[HTML]{EFEFEF}}c 
>{\columncolor[HTML]{EFEFEF}}c @{}}
\toprule
       & \multicolumn{3}{c}{\cellcolor[HTML]{EFEFEF}\textbf{BL 0.001}}                                   & \multicolumn{3}{c}{\textbf{BL 0.01}}                                     & \multicolumn{3}{c}{\cellcolor[HTML]{EFEFEF}\textbf{BL 0.1}}                                         \\ \midrule
       & \multicolumn{2}{c}{\cellcolor[HTML]{EFEFEF}\texttt{LogL}}              & \texttt{KL-qprior}     & \multicolumn{2}{c}{\texttt{LogL}}               & \texttt{KL-qprior}     & \multicolumn{2}{c}{\cellcolor[HTML]{EFEFEF}\texttt{LogL}}                 & \texttt{KL-qprior}      \\ \midrule
Real data   & \multicolumn{2}{c}{\cellcolor[HTML]{EFEFEF}-2484.7627(95.45)}          &                        & \multicolumn{2}{c}{-9019.0723(571.12)}          &                        & \multicolumn{2}{c}{\cellcolor[HTML]{EFEFEF}-43329.2734(2280.71)}          &                         \\
\texttt{VP-FPD} & \multicolumn{2}{c}{\cellcolor[HTML]{EFEFEF}-2518.1560(93.18)}          & 414.8906(3.89)         & \multicolumn{2}{c}{-9025.1680(570.85)}          & 304.8004(5.69)         & \multicolumn{2}{c}{\cellcolor[HTML]{EFEFEF}-43344.3680(2281.89)}          & 246.7808(5.41)          \\
\texttt{VP-LPD} & \multicolumn{2}{c}{\cellcolor[HTML]{EFEFEF}\textbf{-2451.9842(95.00)}} & 29.2950(1.59)          & \multicolumn{2}{c}{\textbf{-9006.1580(571.88)}} & 85.6934(3.81)          & \multicolumn{2}{c}{\cellcolor[HTML]{EFEFEF}\textbf{-43340.0960(2281.90)}} & 180.7204(3.85)          \\
\texttt{VP-NPD} & \multicolumn{2}{c}{\cellcolor[HTML]{EFEFEF}-2454.8502(95.16)}          & \textbf{27.5081(3.67)} & \multicolumn{2}{c}{-9007.7260(571.90)}          & \textbf{85.3090(3.92)} & \multicolumn{2}{c}{\cellcolor[HTML]{EFEFEF}-43340.7120(2281.94)}          & \textbf{180.6803(3.74)} \\ \cmidrule(l){2-10} 
       & \texttt{TL}                        & \texttt{Dist}                     & \texttt{Corr}          & \texttt{TL}            & \texttt{Dist}          & \texttt{Corr}          & \texttt{TL}                           & \texttt{Dist}                     & \texttt{Corr}           \\ \cmidrule(l){2-10} 
Real data   & 0.1239(0.01)                       &                                   &                        & 1.2390(0.11)           &                        &                        & 12.3896(1.09)                         &                                   &                         \\
\texttt{VP-FPD} & 0.2528(0.02)                       & 0.0240(0.00)                      & 0.4636(0.10)           & 1.3463(0.12)           & 0.0506(0.01)           & 0.8966(0.03)           & \textbf{12.4283(1.10)}                & \textbf{0.1773(0.02)}             & 0.9806(0.00)            \\
\texttt{VP-LPD} & \textbf{0.1407(0.02)}              & 0.0152(0.00)                      & \textbf{0.5783(0.09)}  & \textbf{1.2356(0.12)}  & \textbf{0.0479(0.01)}  & \textbf{0.9058(0.02)}  & 12.4716(1.12)                         & 0.1790(0.02)                      & \textbf{0.9810(0.00)}   \\
\texttt{VP-NPD} & 0.1418(0.02)                       & \textbf{0.0151(0.00)}             & 0.5732(0.09)           & 1.2360(0.12)           & \textbf{0.0479(0.01)}  & 0.9055(0.02)           & 12.4633(1.13)                         & 0.1791(0.02)                      & \textbf{0.9810(0.00)}   \\ \bottomrule
\end{tabular}

\end{adjustbox}
\label{tab:bl_all}
\end{table}
First, we assessed the accuracy of branch lengths estimation using datasets generated with distinct \textit{a priori} means over the branch lengths.
Table \ref{tab:bl_all} shows the results of \texttt{VP} models tested on datasets with branch lengths drawn with i.i.d. exponential prior means (rates) of 0.001 (1000), 0.01 (100) and 0.1 (10).
We used the JC69 substitution model in sequence alignment simulation and variational inference models to avoid any effect of evolutionary model parameters on estimating the branch lengths.
For inference, each \texttt{VP} model applies i.i.d. exponential priors on branch lengths. 
The \texttt{VP} model with fixed prior density, \texttt{VP-FPD}, places an i.i.d. prior with mean 0.1, which is usually used by default in variational phylogenetic software \cite{Zhang2019,Fourment2019} for branch lengths and corresponds to the prior used to simulate the third dataset in the current experiment.
We notice in Table \ref{tab:bl_all}, in the case of branch lengths simulated with a prior mean of 0.001, which is severely divergent from the default prior mean of 0.1, that the \texttt{VP-FPD} model overestimates the total tree length (\texttt{TL}) up to 2 times its actual value.
Interestingly, \texttt{VP-L/NPD} models approximate better the branch lengths with smaller average Euclidean distances ($p = 0$, Kruskal-Wallis test), and estimate the \texttt{TL} average values with ratios less than 1.14 to the real \texttt{TL}.
However, when the branch lengths follow distributions close to the default prior, \texttt{VP-FPD} has better branch length estimations ($p = 0$, Kruskal-Wallis test on Euclidean distances). Nonetheless, \texttt{VP-L/NPD} models have fairly similar results, and their estimations are better correlated with the real values ($p = 0$, Kruskal-Wallis test).
Regardless of dataset peculiarities, \texttt{VP-LPD} approximates better the log likelihood (\texttt{LogL}) compared to the other models.
Although all models approximate, up to a point, the actual \texttt{LogL} of the datasets, \texttt{VP-L/NPD} models have smaller \texttt{KL} divergence between the approximate density and prior density (\texttt{KL-qprior}).

Then, we analyzed the consistency of the \texttt{VP} models and the effect of the dataset size on the estimation of branch lengths.
Table \ref{tab:bl_data} shows the results of the estimations on datasets simulated with sequence lengths of 1000 and 5000 bp, number of taxa of 16 and 64, and branch lengths sampled from i.i.d exponential priors of mean 0.001.
For all \texttt{VP} models, the average distances and correlations between the estimate and the actual branch lengths improve with longer sequence alignments but not always with larger trees.
Moreover, the models more accurately estimate the total \texttt{TL} with bigger datasets. 
The improvement is clearly noticed with the \texttt{VP-FPD} model, where its ratio of the estimate and actual \texttt{TL} decreases from 2.11 with the 1000/16 sequence alignment to 1.19 with the 5000/64 sequence alignment.
The \texttt{TL} ratios of the \texttt{VP-L/NPD} models decrease from 1.2 to 1.02 with the same sequence alignments, respectively.
In terms of these performance metrics, \texttt{VP-L/NPD} models have better estimates compared to those of \texttt{VP-FPD}, regardless of the dataset size ($p \le 2.51\mathrm{E}{-13}$, Kruskal-Wallis tests).
Also, for all dataset sizes, the \texttt{VP-LPD} model optimizes the \texttt{LogL} better, and \texttt{VP-L/NPD} models have smaller \texttt{KL-qprior} divergences than those of the \texttt{VP-FPD} model.
\begin{table}[]
\centering
\caption{\small \textbf{Performance of branch lengths (BL) estimation on datasets simulated with different sizes}. The datasets are simulated with JC69 substitution model. BL values are drawn from an exponential distribution with mean \textbf{0.001}. \texttt{VP} models implement a JC69 model and apply an i.i.d. exponential prior on BL. \texttt{VP-FPD} applies a prior with mean 0.1.}
\begin{adjustbox}{width=1\textwidth}
\begin{tabular}{@{}cccccccc@{}}
\toprule
N                      & M                                            & Model                             & \texttt{LogL}                                       & \texttt{KL-qprior}                             & \texttt{TL}                                   & \texttt{Dist}                                 & \texttt{Corr}                                 \\ \midrule
                       & \cellcolor[HTML]{EFEFEF}                     & \cellcolor[HTML]{EFEFEF}Real data & \cellcolor[HTML]{EFEFEF}-1670.4626(52.86)           & \cellcolor[HTML]{EFEFEF}                       & \cellcolor[HTML]{EFEFEF}0.0296(0.01)          & \cellcolor[HTML]{EFEFEF}                      & \cellcolor[HTML]{EFEFEF}                      \\
                       & \cellcolor[HTML]{EFEFEF}                     & \cellcolor[HTML]{EFEFEF}\texttt{VP-FPD}    & \cellcolor[HTML]{EFEFEF}-1678.7043(51.34)           & \cellcolor[HTML]{EFEFEF}94.7330(2.13)          & \cellcolor[HTML]{EFEFEF}0.0625(0.01)          & \cellcolor[HTML]{EFEFEF}0.0117(0.00)          & \cellcolor[HTML]{EFEFEF}0.4852(0.18)          \\
                       & \cellcolor[HTML]{EFEFEF}                     & \cellcolor[HTML]{EFEFEF}\texttt{VP-LPD}    & \cellcolor[HTML]{EFEFEF}\textbf{-1664.4830(52.45)}  & \cellcolor[HTML]{EFEFEF}7.1037(0.82)           & \cellcolor[HTML]{EFEFEF}\textbf{0.0362(0.01)} & \cellcolor[HTML]{EFEFEF}0.0074(0.00)          & \cellcolor[HTML]{EFEFEF}\textbf{0.5925(0.16)} \\
                       & \multirow{-4}{*}{\cellcolor[HTML]{EFEFEF}16} & \cellcolor[HTML]{EFEFEF}\texttt{VP-NPD}    & \cellcolor[HTML]{EFEFEF}-1664.9436(52.45)           & \cellcolor[HTML]{EFEFEF}\textbf{6.4874(0.77)}  & \cellcolor[HTML]{EFEFEF}\textbf{0.0362(0.01)} & \cellcolor[HTML]{EFEFEF}\textbf{0.0073(0.00)} & \cellcolor[HTML]{EFEFEF}0.5882(0.16)          \\ \cmidrule(l){2-8} 
                       &                                              & Real data                         & -2484.7627(95.45)                                   &                                                & 0.1239(0.01)                                  &                                               &                                               \\
                       &                                              & \texttt{VP-FPD}                            & -2518.1560(93.18)                                   & 414.8906(3.89)                                 & 0.2528(0.02)                                  & 0.0240(0.00)                                  & 0.4636(0.10)                                  \\
                       &                                              & \texttt{VP-LPD}                            & \textbf{-2451.9842(95.00)}                          & 29.2950(1.59)                                  & \textbf{0.1407(0.02)}                         & 0.0152(0.00)                                  & \textbf{0.5783(0.09)}                         \\
\multirow{-8}{*}{1000} & \multirow{-4}{*}{64}                         & \texttt{VP-NPD}                            & -2454.8502(95.16)                                   & \textbf{27.5081(3.67)}                         & 0.1418(0.02)                                  & \textbf{0.0151(0.00)}                         & 0.5732(0.09)                                  \\ \midrule
                       & \cellcolor[HTML]{EFEFEF}                     & \cellcolor[HTML]{EFEFEF}Real data & \cellcolor[HTML]{EFEFEF}-8196.2695(193.34)          & \cellcolor[HTML]{EFEFEF}                       & \cellcolor[HTML]{EFEFEF}0.0296(0.01)          & \cellcolor[HTML]{EFEFEF}                      & \cellcolor[HTML]{EFEFEF}                      \\
                       & \cellcolor[HTML]{EFEFEF}                     & \cellcolor[HTML]{EFEFEF}\texttt{VP-FPD}    & \cellcolor[HTML]{EFEFEF}-8196.1490(192.97)          & \cellcolor[HTML]{EFEFEF}123.7540(2.51)         & \cellcolor[HTML]{EFEFEF}0.0361(0.01)          & \cellcolor[HTML]{EFEFEF}0.0039(0.00)          & \cellcolor[HTML]{EFEFEF}0.7871(0.09)          \\
                       & \cellcolor[HTML]{EFEFEF}                     & \cellcolor[HTML]{EFEFEF}\texttt{VP-LPD}    & \cellcolor[HTML]{EFEFEF}\textbf{-8189.3350(193.73)} & \cellcolor[HTML]{EFEFEF}16.8418(1.09)          & \cellcolor[HTML]{EFEFEF}\textbf{0.0310(0.01)} & \cellcolor[HTML]{EFEFEF}\textbf{0.0035(0.00)} & \cellcolor[HTML]{EFEFEF}\textbf{0.8042(0.08)} \\
                       & \multirow{-4}{*}{\cellcolor[HTML]{EFEFEF}16} & \cellcolor[HTML]{EFEFEF}\texttt{VP-NPD}    & \cellcolor[HTML]{EFEFEF}-8189.5360(193.71)          & \cellcolor[HTML]{EFEFEF}\textbf{16.5364(1.12)} & \cellcolor[HTML]{EFEFEF}\textbf{0.0310(0.01)} & \cellcolor[HTML]{EFEFEF}\textbf{0.0035(0.00)} & \cellcolor[HTML]{EFEFEF}0.8041(0.09)          \\ \cmidrule(l){2-8} 
                       &                                              & Real data                         & -12074.5166(440.76)                                 &                                                & 0.1239(0.01)                                  &                                               &                                               \\
                       &                                              & \texttt{VP-FPD}                            & -12089.3860(439.94)                                 & 536.9764(5.30)                                 & 0.1482(0.01)                                  & 0.0076(0.00)                                  & 0.8231(0.04)                                  \\
                       &                                              & \texttt{VP-LPD}                            & \textbf{-12059.1640(441.47)}                        & \textbf{70.4068(2.46)}                         & \textbf{0.1261(0.01)}                         & \textbf{0.0067(0.00)}                         & \textbf{0.8404(0.04)}                         \\
\multirow{-8}{*}{5000} & \multirow{-4}{*}{64}                         & \texttt{VP-NPD}                            & -12061.0200(441.25)                                 & 71.1167(7.69)                                  & 0.1267(0.01)                                  & \textbf{0.0067(0.00)}                         & 0.8393(0.04)                                  \\ \bottomrule
\end{tabular}
\end{adjustbox}
\label{tab:bl_data}
\end{table}

Next, we investigated how the \texttt{VP} models perform on datasets whose trees have external and internal branch lengths drawn from distributions with different means. 
We entertained two scenarios to build such trees (see Table \ref{tab:bl_ext_int}). 
In the first one, we simulated trees with short external branches and longer internal branches (using prior means of 0.005 and 0.1, respectively). 
The second scenario simulates trees with long external branches and shorter internal branches (using prior means of 0.1 and 0.005, respectively).
Besides this branch length distribution detail, we used the same settings for dataset simulation and model inference as in the previous experiment.
\begin{table}[t]
\centering
\caption{\small \textbf{Performance of branch lengths (BL) estimation on datasets simulated with external (ext) and internal (int) BL having different prior means}. The datasets have 64 sequences of length 1000 bp and are simulated with JC69 substitution model. BL values are drawn from an exponential distribution. \texttt{VP} models implement a JC69 model and apply an i.i.d. exponential prior on BL. \texttt{VP-FPD} applies a prior with mean 0.1.}
\begin{adjustbox}{width=1\textwidth}
\begin{tabular}{@{}c
>{\columncolor[HTML]{EFEFEF}}c 
>{\columncolor[HTML]{EFEFEF}}c 
>{\columncolor[HTML]{EFEFEF}}c 
>{\columncolor[HTML]{EFEFEF}}c 
>{\columncolor[HTML]{EFEFEF}}c 
>{\columncolor[HTML]{EFEFEF}}c cccccc@{}}
\toprule
       & \multicolumn{6}{c}{\cellcolor[HTML]{EFEFEF}\textbf{BL (ext0.005, int0.1)}}                                                                                                                                          & \multicolumn{6}{c}{\textbf{BL (ext0.1, int0.005)}}                                                                                                \\ \midrule
       & \multicolumn{2}{c}{\cellcolor[HTML]{EFEFEF}\texttt{LogL}}                 & \multicolumn{2}{c}{\cellcolor[HTML]{EFEFEF}\texttt{KL-qprior}}      & \multicolumn{2}{c}{\cellcolor[HTML]{EFEFEF}\texttt{TL}}           & \multicolumn{2}{c}{\texttt{LogL}}                 & \multicolumn{2}{c}{\texttt{KL-qprior}}        & \multicolumn{2}{c}{\texttt{TL}}               \\ \cmidrule(l){2-13} 
Real data   & \multicolumn{2}{c}{\cellcolor[HTML]{EFEFEF}-24300.4082(1692.34)}          & \multicolumn{2}{c}{\cellcolor[HTML]{EFEFEF}}                        & \multicolumn{2}{c}{\cellcolor[HTML]{EFEFEF}6.2972(0.73)}          & \multicolumn{2}{c}{-26387.3984(1863.89)}          & \multicolumn{2}{c}{}                          & \multicolumn{2}{c}{6.7119(0.76)}              \\
\texttt{VP-FPD} & \multicolumn{2}{c}{\cellcolor[HTML]{EFEFEF}-24313.7400(1693.06)}          & \multicolumn{2}{c}{\cellcolor[HTML]{EFEFEF}289.6527(5.46)}          & \multicolumn{2}{c}{\cellcolor[HTML]{EFEFEF}6.3608(0.73)}          & \multicolumn{2}{c}{-26400.9060(1864.33)}          & \multicolumn{2}{c}{295.4958(5.73)}            & \multicolumn{2}{c}{6.7690(0.76)}              \\
\texttt{VP-LPD} & \multicolumn{2}{c}{\cellcolor[HTML]{EFEFEF}\textbf{-24296.1740(1693.27)}} & \multicolumn{2}{c}{\cellcolor[HTML]{EFEFEF}119.6626(3.69)}          & \multicolumn{2}{c}{\cellcolor[HTML]{EFEFEF}6.3325(0.74)}          & \multicolumn{2}{c}{\textbf{-26383.8200(1864.33)}} & \multicolumn{2}{c}{128.7518(4.00)}            & \multicolumn{2}{c}{\textbf{6.7296(0.77)}}     \\
\texttt{VP-NPD} & \multicolumn{2}{c}{\cellcolor[HTML]{EFEFEF}-24297.7720(1693.24)}          & \multicolumn{2}{c}{\cellcolor[HTML]{EFEFEF}\textbf{119.4662(3.78)}} & \multicolumn{2}{c}{\cellcolor[HTML]{EFEFEF}\textbf{6.3296(0.75)}} & \multicolumn{2}{c}{-26385.5320(1864.40)}          & \multicolumn{2}{c}{\textbf{128.4644(3.98)}}   & \multicolumn{2}{c}{6.7327(0.78)}              \\ \cmidrule(l){2-13} 
       & \multicolumn{2}{c}{\cellcolor[HTML]{EFEFEF}All branches}                  & \multicolumn{2}{c}{\cellcolor[HTML]{EFEFEF}Externals}               & \multicolumn{2}{c}{\cellcolor[HTML]{EFEFEF}Internals}             & \multicolumn{2}{c}{All branches}                  & \multicolumn{2}{c}{Externals}                 & \multicolumn{2}{c}{Internals}                 \\ \cmidrule(l){2-13} 
       & \texttt{Dist}                        & \texttt{Corr}                      & \texttt{Dist}                     & \texttt{Corr}                   & \texttt{Dist}                    & \texttt{Corr}                  & \texttt{Dist}           & \texttt{Corr}           & \texttt{Dist}         & \texttt{Corr}         & \texttt{Dist}         & \texttt{Corr}         \\ \cmidrule(l){2-13} 
\texttt{VP-FPD} & \textbf{0.1279(0.02)}                & 0.9871(0.00)                       & 0.0330(0.01)                      & 0.7536(0.09)                    & \textbf{0.1246(0.02)}            & 0.9818(0.01)                   & 0.1213(0.02)            & 0.9895(0.00)            & \textbf{0.1180(0.02)} & 0.9854(0.00)          & 0.0318(0.01)          & 0.7413(0.09)          \\
\texttt{VP-LPD} & 0.1281(0.02)                         & \textbf{0.9875(0.00)}              & \textbf{0.0281(0.01)}             & \textbf{0.7981(0.07)}           & 0.1257(0.02)                     & \textbf{0.9821(0.01)}          & \textbf{0.1207(0.02)}   & \textbf{0.9898(0.00)}   & 0.1183(0.02)          & \textbf{0.9857(0.00)} & \textbf{0.0271(0.00)} & \textbf{0.7870(0.08)} \\
\texttt{VP-NPD} & 0.1280(0.02)                         & \textbf{0.9875(0.00)}              & \textbf{0.0281(0.01)}             & 0.7963(0.07)                    & 0.1257(0.02)                     & \textbf{0.9821(0.01)}          & 0.1209(0.02)            & \textbf{0.9898(0.00)}   & 0.1185(0.02)          & \textbf{0.9857(0.00)} & \textbf{0.0271(0.00)} & 0.7847(0.08)          \\ \bottomrule
\end{tabular}
\end{adjustbox}
\label{tab:bl_ext_int}
\end{table}
Overall, the estimation results shown in Table \ref{tab:bl_ext_int} are relatively similar for the three \texttt{VP} models. 
However, we note that \texttt{VP-LPD} model optimizes better the \texttt{LogL}, and \texttt{VP-L/NPD} models have smaller \texttt{KL-qprior} divergences, as found in the previous experiment.
In the first scenario, branch length estimations of the \texttt{VP-FPD} model have slightly smaller average distances with actual values (\(p = 2.63\mathrm{E}{-05}\), Kruskal-Wallis test). The external branches (with means of 0.005) are estimated better with \texttt{VP-L/NPD} models, and the internal branches (with means of 0.1) are estimated better with the \texttt{VP-FPD} model.
The second scenario has an opposite outcome compared to the first one. The \texttt{VP-LPD} model estimates branch lengths with smaller average distances to the actual values (\(p = 4.19\mathrm{E}{-189}\), Kruskal-Wallis test). The external branches (with means of 0.1) are estimated better with \texttt{VP-FPD} model, and the internal branches (with means of 0.005) are estimated better with \texttt{VP-L/NPD} models.
Thus, \texttt{VP-L/NPD} models achieve better estimations of either external or internal branch lengths drawn from distributions different from the default prior.

\subsection{Substitution model parameters estimation performance}
After that, we evaluated the performance of the \texttt{VP} models in approximating the posterior densities of the substitution model parameters.
Usually, the model parameters are global and inferred for the whole sequence alignment. 
Here, we are interested in estimating the parameters of the GTR model, parameterized by a set of six substitution rates and another of four relative frequencies.
We assume that each set of parameters sums to one, so we can use Dirichlet distributions as priors over them.
To simplify the evaluation scenarios for the estimations of the six substitution rates, we used the HKY85 model \cite{Hasegawa1985} to simulate the sequence alignments.
HKY85 is a special case of the GTR model that defines a ratio of transition and transversion rates named \(\bm{\kappa}\) and a set of relative frequencies.
Thus, varying the ratio \(\bm{\kappa}\) will change the values of two class rates at once.
For the other simulation settings, we sampled branch length values from i.i.d. exponential priors of mean 0.1, \(\bm{\kappa}\) values from gamma distributions (default mean equals 1), and the relative frequencies from Dirichlet distributions (default concentrations equal 10 for simulation and 1 for inference).
We implemented the \texttt{VP} models with a GTR model. We applied i.i.d. exponential priors on branch lengths and Dirichlet priors on substitution rates and relative frequencies. The \texttt{VP-FPD} and \texttt{VP-L/NPD} models implement default and adaptable prior parameters, respectively, for all phylogenetic parameters.

\begin{table}[t]
\centering
\caption{\small \textbf{Performance of substitution rates estimation on datasets simulated with different prior means on the transition/transversion rate ratio \(\bm{\kappa}\)}. The datasets have 64 sequences of length 1000 bp and are simulated with HKY85 substitution model. \(\bm{\kappa}\) values are drawn from gamma distribution. Relative frequencies are drawn from Dirichlet distribution with concentrations of 10. BL values are drawn from an exponential distribution of mean 0.1. \texttt{VP} models implement a GTR model and apply an i.i.d. exponential prior on BL and Dirichlet priors on rates and relative frequencies. \texttt{VP-FPD} applies a Dirichlet prior with concentration 1.}
\begin{adjustbox}{width=1\textwidth}
\begin{tabular}{@{}c
>{\columncolor[HTML]{EFEFEF}}c 
>{\columncolor[HTML]{EFEFEF}}c cc
>{\columncolor[HTML]{EFEFEF}}c 
>{\columncolor[HTML]{EFEFEF}}c @{}}
\toprule
       & \multicolumn{2}{c}{\cellcolor[HTML]{EFEFEF}$\bm{\kappa}$ \textbf{0.25}} & \multicolumn{2}{c}{$\bm{\kappa}$ \textbf{1}}            & \multicolumn{2}{c}{\cellcolor[HTML]{EFEFEF}$\bm{\kappa}$ \textbf{4}} \\ \midrule
       & \texttt{LogL}                         & \texttt{KL-qprior}              & \texttt{LogL}                 & \texttt{KL-qprior}      & \texttt{LogL}                        & \texttt{KL-qprior}            \\ \cmidrule(l){2-7} 
Real data   & -41403.2227(2158.37)                  &                                 & -42732.2461(2179.98)          &                         & -40114.5391(2028.43)                 &                               \\
\texttt{VP-FPD} & -41424.0040(2159.88)                  & 270.0602(5.28)                  & -42753.3000(2181.57)          & 270.5125(5.40)          & -40134.8240(2029.26)                 & 267.1696(5.47)                \\
\texttt{VP-LPD} & \textbf{-41418.5160(2159.88)}         & 184.4728(4.15)                  & \textbf{-42747.5040(2181.22)} & 185.6610(4.08)          & \textbf{-40128.5920(2029.27)}        & 181.7221(3.87)                \\
\texttt{VP-NPD} & -41418.7160(2160.08)                  & \textbf{182.2155(4.10)}         & -42747.5360(2181.73)          & \textbf{183.1652(4.08)} & -40128.8320(2029.29)                 & \textbf{179.3205(3.70)}       \\ \cmidrule(l){2-7} 
       & \texttt{Dist}                         & \texttt{Corr}                   & \texttt{Dist}                 & \texttt{Corr}           & \texttt{Dist}                        & \texttt{Corr}                 \\ \cmidrule(l){2-7} 
\texttt{VP-FPD} & 0.0185(0.01)                          & 0.9950(0.00)                    & 0.0190(0.01)                  & 0.5226(0.40)            & 0.0182(0.01)                         & 0.9980(0.00)                  \\
\texttt{VP-LPD} & \textbf{0.0175(0.01)}                 & \textbf{0.9955(0.00)}           & 0.0181(0.01)                  & 0.5360(0.40)            & 0.0176(0.01)                         & \textbf{0.9981(0.00)}         \\
\texttt{VP-NPD} & \textbf{0.0175(0.01)}                 & 0.9954(0.00)                    & \textbf{0.0178(0.01)}         & \textbf{0.5407(0.39)}   & \textbf{0.0173(0.01)}                & \textbf{0.9981(0.00)}         \\ \bottomrule
\end{tabular}
\end{adjustbox}
\label{tab:rates}
\end{table}
\begin{table}[]
\centering
\caption{\small \textbf{Performance of relative frequencies estimation on datasets simulated with different prior means}. The datasets have 64 sequences of length 1000 bp and are simulated with HKY85 substitution model. Relative frequencies values are drawn from Dirichlet distribution with different nucleotide concentrations. For instance, \textbf{Dir(10AT)} means nucleotides \textbf{A} and \textbf{T} have concentrations of 10 and the other two concentrations are 1. \textbf{Dir(10)} means all concentrations equal to 10. \(\bm{\kappa}\) values are drwan from gamma distribution with mean 1. BL values are drawn from an exponential distribution of mean 0.1. \texttt{VP} models implement a GTR model and apply an i.i.d. exponential prior on BL and Dirichlet priors on rates and relative frequencies. \texttt{VP-FPD} applies a Dirichlet prior with concentration 1.}
\begin{adjustbox}{width=1\textwidth}

\begin{tabular}{@{}c
>{\columncolor[HTML]{EFEFEF}}c 
>{\columncolor[HTML]{EFEFEF}}c cc
>{\columncolor[HTML]{EFEFEF}}c 
>{\columncolor[HTML]{EFEFEF}}c @{}}
\toprule
       & \multicolumn{2}{c}{\cellcolor[HTML]{EFEFEF}\textbf{Dir(10AT)}} & \multicolumn{2}{c}{\textbf{Dir(10)}}                    & \multicolumn{2}{c}{\cellcolor[HTML]{EFEFEF}\textbf{Dir(10GC)}} \\ \midrule
       & \texttt{LogL}                     & \texttt{KL-qprior}         & \texttt{LogL}                 & \texttt{KL-qprior}      & \texttt{LogL}                     & \texttt{KL-qprior}         \\ \cmidrule(l){2-7} 
Real data   & -34683.3945(3061.60)              &                            & -42732.2461(2179.98)          &                         & -34382.1719(3514.33)              &                            \\
\texttt{VP-FPD} & -34703.5800(3063.90)              & 255.2194(8.09)             & -42753.2040(2181.45)          & 270.5957(5.53)          & -34401.6440(3515.04)              & 254.0669(8.25)             \\
\texttt{VP-LPD} & -34696.9040(3064.24)              & 172.3171(5.89)             & \textbf{-42747.5880(2181.26)} & 185.5565(4.02)          & \textbf{-34394.8080(3514.94)}     & 172.0875(6.25)             \\
\texttt{VP-NPD} & \textbf{-34696.7960(3064.05)}     & \textbf{170.4964(5.66)}    & -42747.6960(2181.33)          & \textbf{183.1414(4.00)} & -34394.8920(3515.20)              & \textbf{170.3758(6.09)}    \\ \cmidrule(l){2-7} 
       & \texttt{Dist}                     & \texttt{Corr}              & \texttt{Dist}                 & \texttt{Corr}           & \texttt{Dist}                     & \texttt{Corr}              \\ \cmidrule(l){2-7} 
\texttt{VP-FPD} & 0.0130(0.01)                      & 0.9994(0.00)               & 0.0147(0.01)                  & 0.9899(0.02)            & 0.0123(0.01)                      & 0.9995(0.00)               \\
\texttt{VP-LPD} & 0.0127(0.01)                      & 0.9994(0.00)               & 0.0143(0.01)                  & 0.9907(0.02)            & 0.0120(0.01)                      & 0.9995(0.00)               \\
\texttt{VP-NPD} & \textbf{0.0123(0.01)}             & \textbf{0.9995(0.00)}      & \textbf{0.0140(0.01)}         & \textbf{0.9911(0.02)}   & \textbf{0.0119(0.01)}             & 0.9995(0.00)               \\ \bottomrule
\end{tabular}
\end{adjustbox}
\label{tab:freqs}
\end{table}

In Table \ref{tab:rates}, we present the performance of substitution rates estimation on three datasets simulated with ratios \(\bm{\kappa}\) drawn with prior means of 0.25, 1 and 4.
We noted that the \texttt{VP-L/NPD} models estimate the substitution rates with smaller average distances and better average correlations with actual rates ($p = 0$, Kruskal-Wallis tests) more accurately than those of the \texttt{VP-FPD} model, even, surprisingly, for the dataset simulated with default prior means. 
Moreover, they optimize better the \texttt{LogL} of the three datasets, and their \texttt{KL-qprior} divergences are smaller compared to the default model.
Regarding evaluating the estimation of the relative frequencies, we simulated sequence alignments with different nucleotide content prior distributions. 
Table \ref{tab:freqs} highlights the results of the \texttt{VP} models applied to three datasets characterized by an AT-rich, equally distributed and GC-rich nucleotide content, respectively.
As for the estimation of substitution rates, \texttt{VP-L/NPD} models perform well in estimating the relative frequencies and the \texttt{LogL} of the data. Further, the \texttt{VP-NPD} model has the best estimations for the three datasets compared to the others ($p = 0$, Kruskal-Wallis tests).

\subsection{Convergence analysis}
\begin{figure*}[]
    \begin{subfigure}[]{1\textwidth}
    \centering
    \caption{\small BL simulated with a prior mean 0.001}
    \includegraphics[width=\textwidth, trim={0 0.9cm 0 0}, clip]{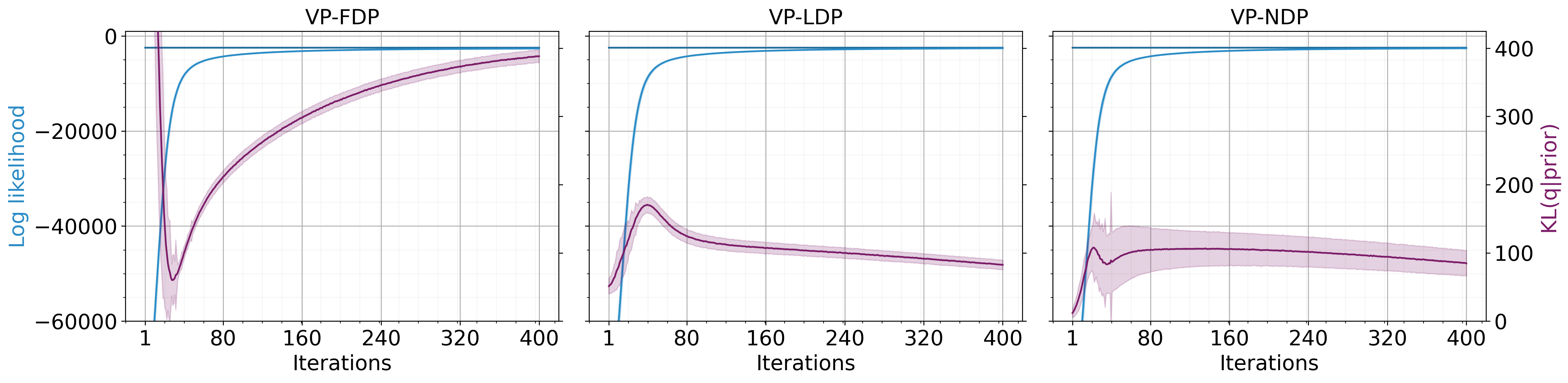}
    \end{subfigure}

    \begin{subfigure}[]{1\textwidth}
    \centering
    \includegraphics[width=0.94\textwidth, trim={0 0 0 1cm}, clip]{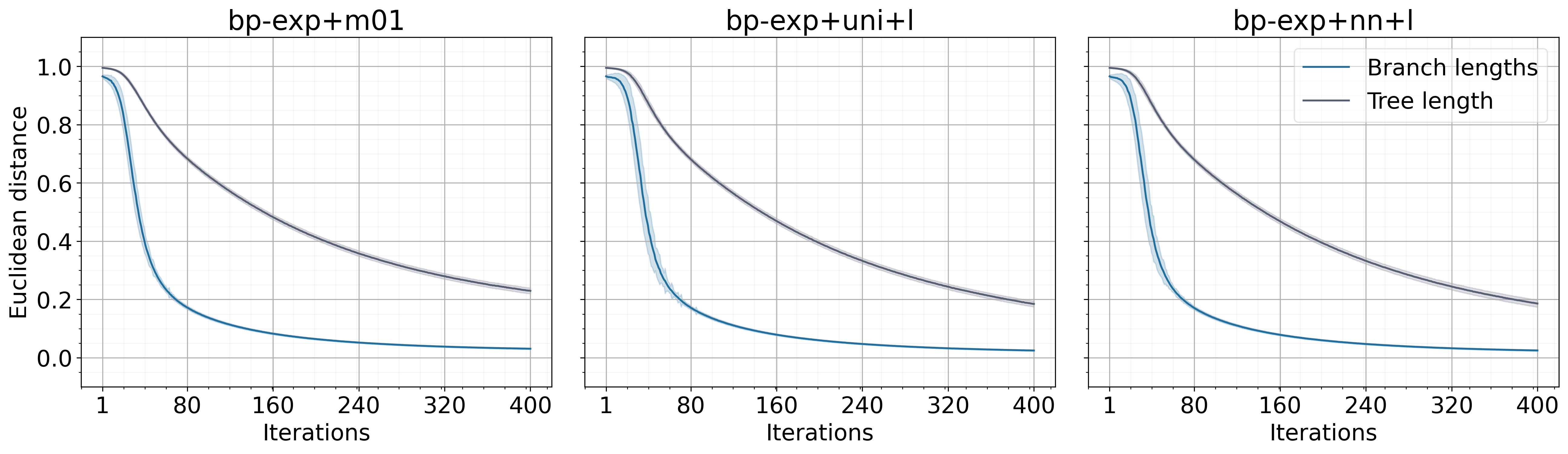}
    \end{subfigure}

    \begin{subfigure}[]{1\textwidth}
    \centering
    \caption{\small BL simulated with a prior mean 0.01}
    \includegraphics[width=\textwidth, trim={0 0.9cm 0 0}, clip]{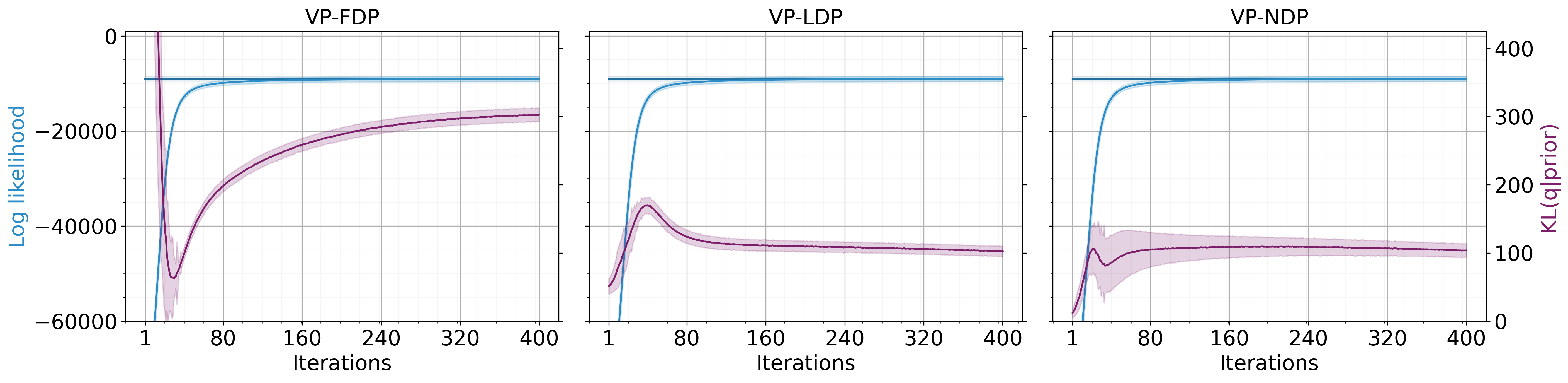}
    \end{subfigure}

    \begin{subfigure}[]{1\textwidth}
    \centering
    \includegraphics[width=0.94\textwidth, trim={0 0 0 1cm}, clip]{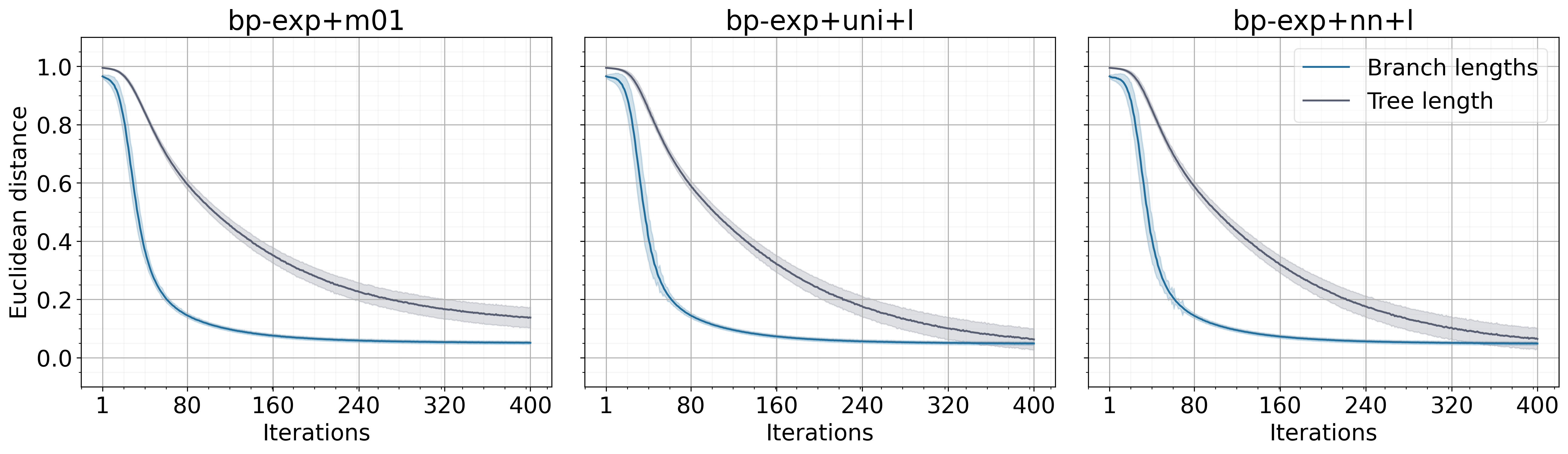}
    \end{subfigure}

    \begin{subfigure}[]{1\textwidth}
    \centering
    \caption{\small BL simulated with a prior mean 0.1}
    \includegraphics[width=\textwidth, trim={0 0.9cm 0 0}, clip]{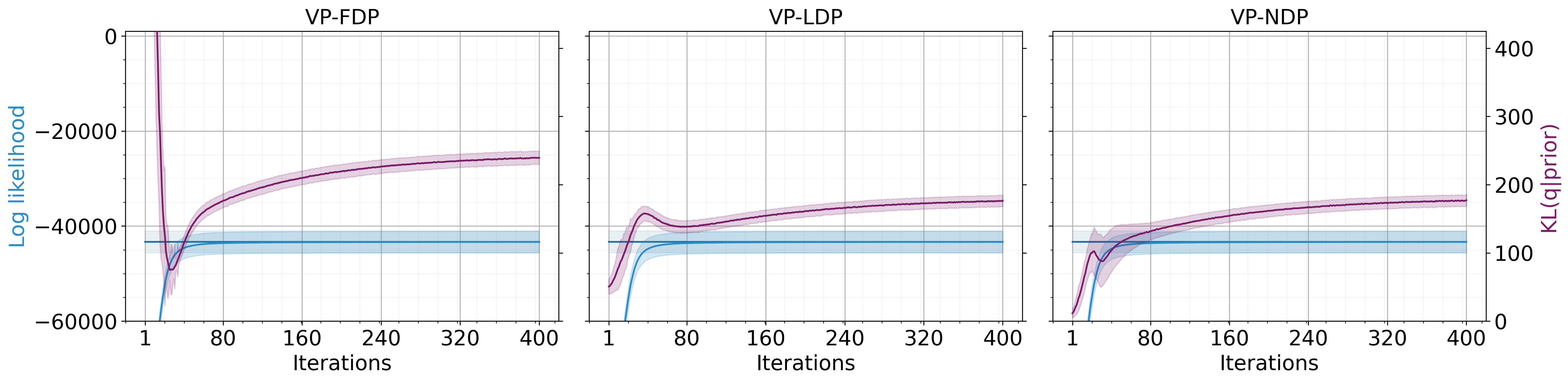}
    \end{subfigure}

    \begin{subfigure}[]{1\textwidth}
    \centering
    \includegraphics[width=0.94\textwidth, trim={0 0 0 1cm}, clip]{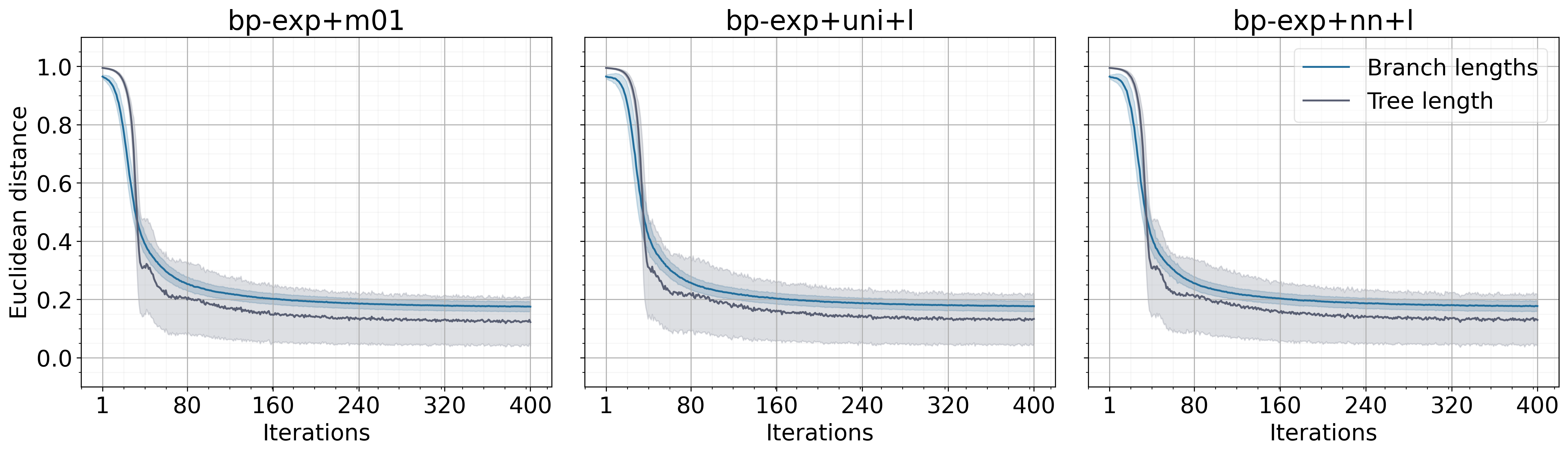}
    \end{subfigure}

    \caption{\small \textbf{Convergence and performance of the \texttt{VP} models for branch length (BL) estimation on datasets simulated with different prior means}. The datasets have 64 sequences of length 1000 bp and are simulated with JC69 substitution model. BL values are drawn from an exponential distribution. \texttt{VP} models implement a JC69 model and apply an i.i.d. exponential prior on BL. \texttt{VP-FPD} applies a prior with mean 0.1. The first 400 training iterations are shown.} 
    \label{fig:conv}
\end{figure*}

Finally, we studied the convergence progression of the \texttt{VP} models when fitted on different datasets generated with three prior means (0.001, 0.01 and 0.1) for branch lengths. 
Figure \ref{fig:conv} illustrates the performance results during the training step using the models' \texttt{LogL} and the \texttt{KL-qprior} divergences. 
We also report the scaled Euclidean distances of the branch length estimates with their actual values.
The figure shows the averages and standard deviations of the metrics, which were calculated over 100 data times ten fit replicates.

First, in all the datasets, the \texttt{LogL} of the three \texttt{VP} models converges to the actual average \texttt{LogL} within the first 400 iterations.
However, the \texttt{LogL} convergence is slower in datasets with shorter branches than in those with longer ones.

Second, The convergence trend of the \texttt{KL-qprior} divergences is most particular, as shown in Figure \ref{fig:conv}.
We noticed that the \texttt{KL-qprior} divergences of the two types of \texttt{VP} models have different progression trajectories during their training across all the datasets.
On the one hand, optimizing the \texttt{VP-FPD} models starts with large \texttt{KL-qprior} values (\(> 1000\)). Then, the \texttt{KL-qprior} values drop sharply (\(\approx 100\)) around the fortieth iteration as the \texttt{LogL} estimations start to converge. After the drop, they return to increase to higher values gradually.
On the other hand, the optimization of the \texttt{VP-L/NPD} models starts with small \texttt{KL-qprior} values (\(< 100\)). They increase slowly as the \texttt{LogL} values start to converge until the fortieth iteration (\(< 200\)). After that, they return to decrease gradually for datasets with shorter branches, but they continue to increase for datasets with longer branches.

Last, for each dataset, the branch length Euclidean distances of the three models have a similar convergence progression. 
As with the \texttt{LogL}, the convergence of branch length distances to small values is slower in datasets with shorter branches than in those with longer branches. This trend is more noticeable with the total \texttt{TL}. 
However, when we compare the convergence of the \texttt{LogL} estimations and the branch length distances, we find that the \texttt{LogL} estimations converge faster than the distances. This finding suggests that the \texttt{LogL} has multiple local and nearby maxima that can be reached with distinct branch length estimates that are not close to the actual values.

\begin{table}[]
\centering
\small
\caption{\small \textbf{Running times of the \texttt{VP} models for branch length (BL) estimation on datasets simulated with different prior means}. The times are reported in seconds. The datasets have 64 sequences of length 1000 bp and are simulated with JC69 substitution model. BL values are drawn from an exponential distribution. \texttt{VP} models implement a JC69 model and apply an i.i.d. exponential prior on BL. \texttt{VP-FPD} applies a prior with mean 0.1.}

\begin{tabularx}{\linewidth}{c
>{\columncolor[HTML]{EFEFEF}}Y Y
>{\columncolor[HTML]{EFEFEF}}Y}
\toprule
\multicolumn{1}{l}{} & \textbf{BL 0.001}      & \textbf{BL 0.01}        & \textbf{BL 0.1}        \\ \midrule
\texttt{VP-FPD}      & \textbf{83.1292(2.37)} & 96.5395(17.69)          & 102.8235(23.20)        \\
\texttt{VP-LPD}      & 84.4600(1.66)          & \textbf{93.1648(12.06)} & 101.0446(19.15)        \\
\texttt{VP-NPD}      & 83.1529(2.01)          & 96.6440(1.46)           & \textbf{97.4801(7.55)} \\ \bottomrule
\end{tabularx}

\label{tab:bl_times}
\end{table}
In addition, we measured the empirical running times required to fit the \texttt{VP} models during 2000 iterations (see Table \ref{tab:bl_times}). 
For all the datasets, the running times of the three models are close and of the same order of magnitude. However, the \texttt{VP-L/NPD} models need slightly less time to fit datasets with more divergent sequences.
Regardless of the models, the running times increase largely with the increase of the number of taxa and moderately with the increase of the alignment length (results not shown).

\section{Discussion}
Recent applications of variational Bayesian phylogeny have assigned default prior distributions to the likelihood parameters \cite{Fourment2019,Zhang2019,Fourment2020}, which can lead to biased and excessively high posterior probabilities \cite{Huelsenbeck2002,Nascimento2017}.
Here, we demonstrated that variational phylogenetic (\texttt{VP}) models using misspecified prior densities are prone to bias when the data are weak.
For example, we showed that a \texttt{VP} model estimates twice the total length of the tree when using default and independent exponential priors on branch lengths with relatively similar and short sequences.
This finding is supported by several MCMC-based studies that have analyzed and explained the effect of branch length priors on the posterior resulting in very long trees \cite{Yang2005a,Kolaczkowski2007,Brown2010,Fabreti2022}.
However, we found that the estimation of the substitution model parameters using \texttt{VP} models with default priors is less biased, even when the actual parameters are far from the range of the default priors. 
It was reported that sequence evolutionary parameters are relatively insensitive to the prior choice when estimated using the whole sequence alignment and less complex models \cite{Alfaro2006}.

In this paper, we introduced a variational phylogenetic approach that provides flexibility to the prior densities, making it insensitive to inappropriate prior specifications.
Using a gradient ascent strategy, the approach implements adaptable parameters for the prior distributions that will be jointly learned from the data with the variational approximate parameters.
The prior learning in variational inference (VI) is connected to the Empirical Bayes (EB), where EB and VI estimate the prior parameters by maximizing the marginal likelihood and the evidence lower bound, respectively \cite{Tomczak2018,Fortuin2022}.
We implemented two \texttt{VP} models with adaptable prior densities, \texttt{VP-LPD} and \texttt{VP-NPD}, using random uniform-sampled and neural networks-generated initializations, respectively.
We showed that regardless of the type of initialization, the models perform better in estimating phylogenetic parameters than \texttt{VP} models with fixed priors that differ from the actual values of the parameters.
Moreover, the \texttt{VP-NPD} model improved the accuracy of the estimation of the sequence evolutionary parameters. However, its accuracy decreases in estimating the vector of branch lengths.
Furthermore, an advantage of using a neural network-based prior parameterization is reducing the burning step (in the first fortieth iterations) and speeding up the model convergence (Figure \ref{fig:conv}). 

This work is the first step for implementing and evaluating the \texttt{VP} models using adaptable prior densities with different parameterization strategies. 
Nevertheless, it has some limitations that can be addressed in future work. 
The current design of prior models does not capture the correlations between the branch lengths nor between phylogenetic parameters.
More sophisticated prior models could be evaluated using suitable simulation-based scenarios, such as compound Dirichlet priors for branch lengths \cite{Zhang2012a}, coalescent priors for rooted trees \cite{Fourment2019,Zhang2022,Ki2022} and rate heterogeneity among sites \cite{Yang1996,Fourment2019}.
Moreover, we hypothesize that using a proper neural network architecture for prior and variational parameterizations could help capture such complex correlations.
Considering that our variational models estimate the phylogenetic parameters given a fixed tree, it would be interesting to investigate how we could implement adaptable prior models with a fully Bayesian phylogenetic inference approach such as \texttt{VBPI} \cite{Zhang2022}, which applies a uniform prior on the tree topology.
\texttt{VBPI} represents the variational posterior of the tree topology with a flexible distribution (subsplit Bayesian networks \cite{Zhang2018a}) and a structured amortization of the branch lengths over tree topologies, which could make implementing the adaptable prior models a stimulating challenge.
Last but not least, we envision assessing the performance and convergence of the \texttt{VP} models with empirical datasets as done in previous variational phylogenetic applications \cite{Zhang2019,Fourment2019,Dang2019,Ki2022}.

%
%
%
\bibliographystyle{splncs04}
\bibliography{2023_remita_recombcg}
\end{document}